\documentclass[11pt,twoside]{article}

\usepackage{asp2006}
\usepackage{epsf}
\usepackage{psfig}
\usepackage{lscape}

\begin{document}
\title{VIMOS spectroscopy of extended emission line regions}   
\author{K. J. Inskip$^1$, M. Villar-Martin$^2$, C. Tadhunter$^1$, R. Morganti$^3$,
  J. Holt$^1$, D. Dicken$^1$}   
\affil{$^1$Department of Physics \& Astronomy, University of Sheffield, Sheffield S3
 7RH\\$^2$Instituto de Astrof\'{i}sica de Andaluc\'{i}a (CSIC), Aptdo. 3004,
 18080 Granada, Spain\\$^3$Netherlands Foundation for Research in Astronomy, Postbus 2, 79
90 AA Dwingeloo, the Netherlands}    

\begin{abstract} 
We present the results of spectroscopic and imaging observations of
the FRII radio galaxies PKS2250-41 and PKS1932-46. 
Both sources display very extensive emission line regions, and  appear to
be undergoing interactions with companion bodies.   In addition to
disturbed gas kinematics associated with interactions with the radio
source, the more distant
emitting material displays simple, narrow emission line
profiles, often at significant velocity offsets from the system
rest-frame, and may be associated with tidal debris.

\end{abstract}



\section{Introduction}

Extended emission line regions (EELRs) are often observed around
powerful distant radio sources \citep{mcc87}, and their observed
properties (size, luminosity, kinematics and ionization state) depend
strongly on those of the radio source
\citep[e.g.][]{best00,inskip02a,mrv02}. While the behaviour of the
emitting material and balance between different ionization mechanisms
is becoming increasingly well understood, there are still many
outstanding questions.  Contrary to expectation, the emitting material
is not always limited to lying along the  radio source axis or within
AGN ionization cones. Recent narrow band imaging observations
\citep{vm05,si02,tad00} have  identified extensive emission line
regions lying almost perpendicular to the radio axis, which may
perhaps be related to the large gaseous halos observed at
higher redshifts.   The origin of the extended
emission line gas  -- whether it exists {\it in situ} prior to the
onset of the radio source, is formed from material driven out of  the
host galaxy by AGN/radio source/starburst related outflows, or is
produced via galaxy interactions/mergers associated with the radio
source triggering event -- is an interesting problem, and one which
may be linked with both the origin of radio source activity and the
host galaxy evolution.

Integral field spectroscopy (IFS) has the potential to greatly improve
our understanding of EELRs, particularly in terms of building
up a coherent picture of the distribution, nature and
origin of the emission line gas, and the links with the radio source
triggering mechanism.  We have carried out IFS observations of PKS1932-46 and PKS2250-41,
using  VIMOS \citep{lf03,s05} on the VLT UT3 during
2004/2005.  The observations were reduced using \textsc{vipgi}
\citep{z05,s05}; full details are available in \citet{inskip06}.  
Additionally, long-slit spectroscopic observations of PKS2250-41 at a
PA of $69^{\circ}$ were obtained  
using FORS1 at the VLT UT2, and reduced using standard IRAF routines.
We also include the results of Spitzer MIPS photometry of PKS1932-46 \citep{dicken07}.

\section{Results}

\begin{figure*}
\vspace{1.6 in}
\begin{center}
\includegraphics{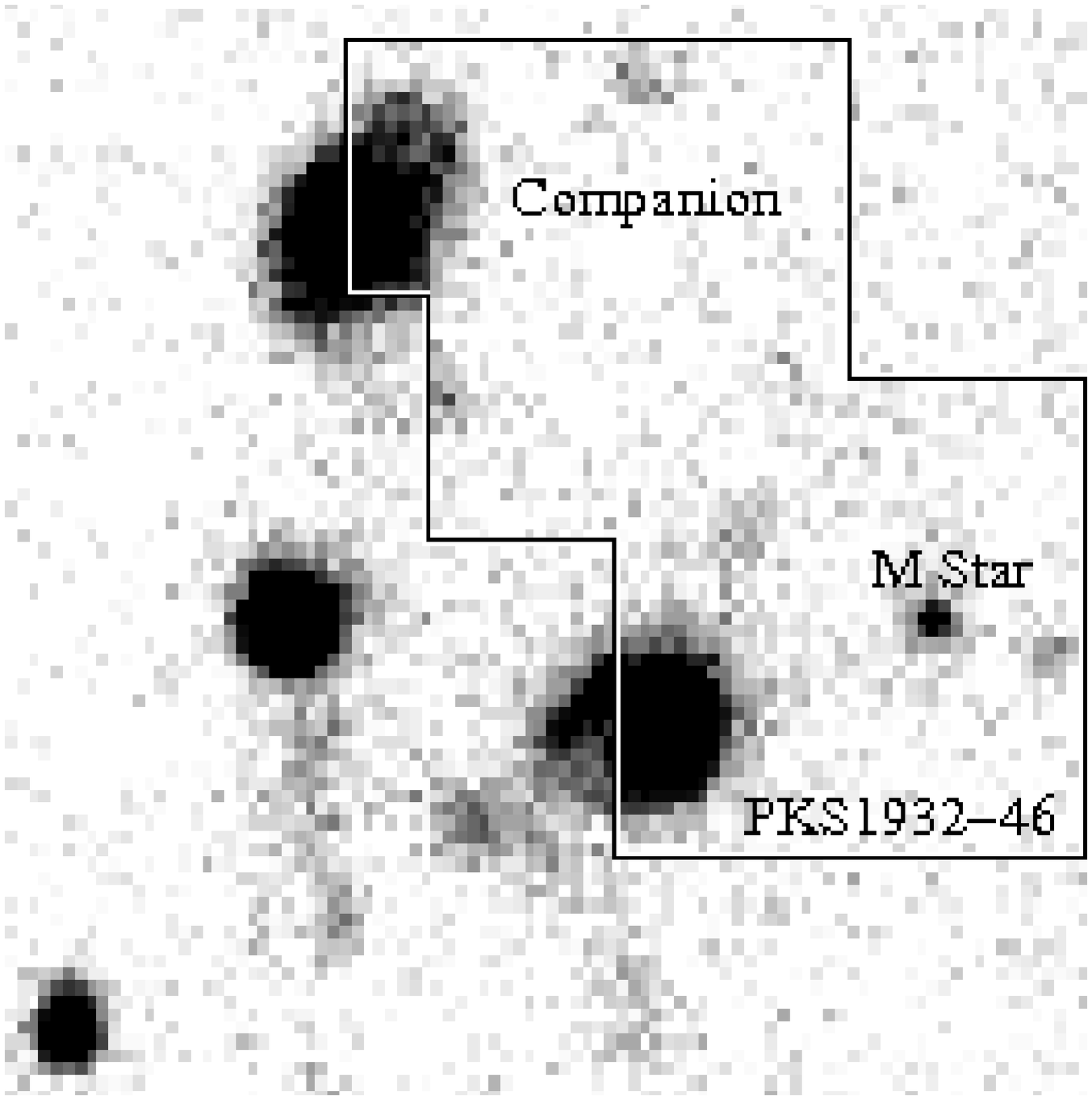}
\includegraphics{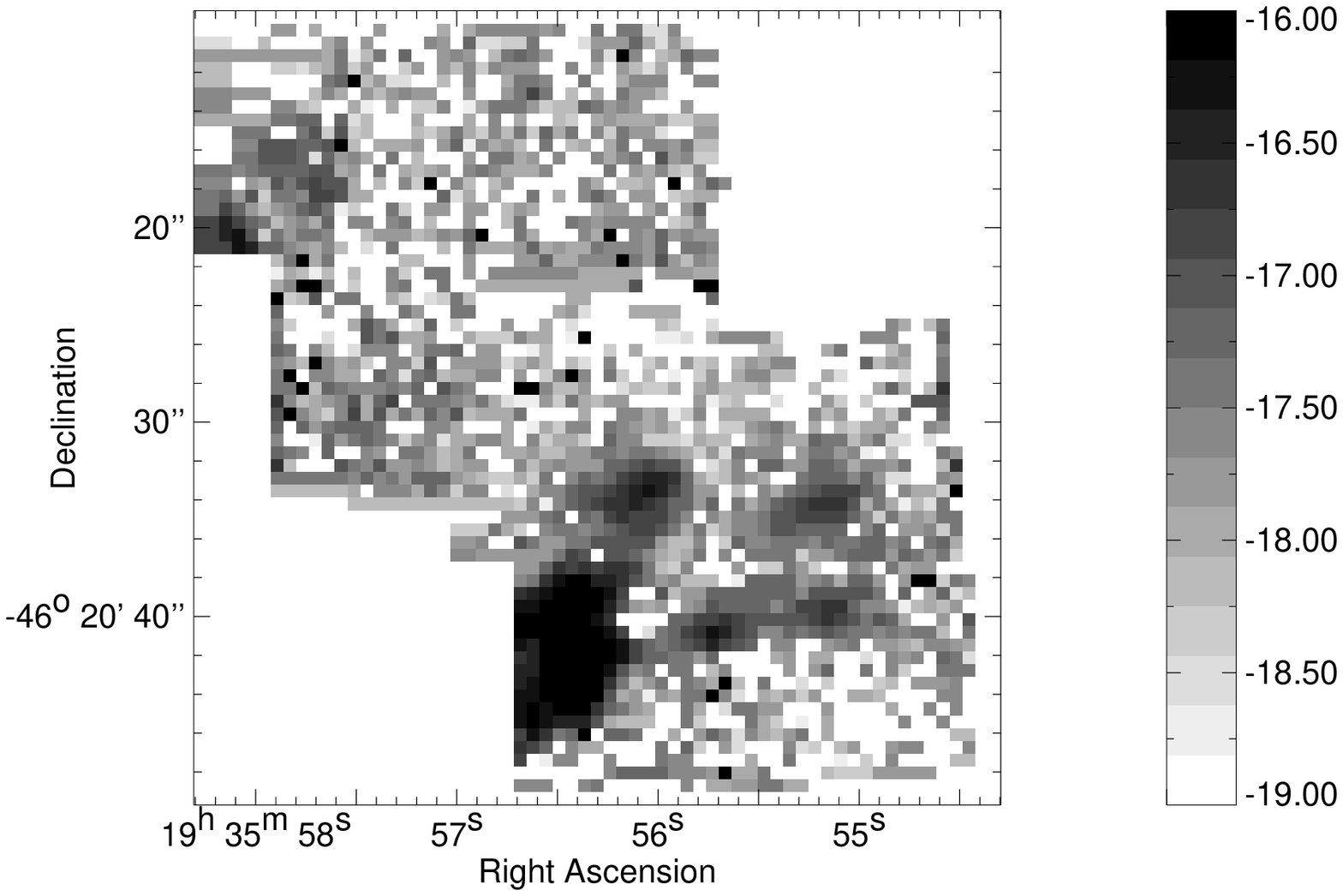}
\includegraphics{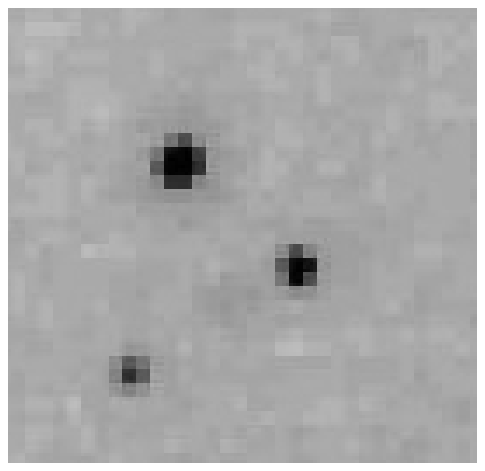}
\includegraphics{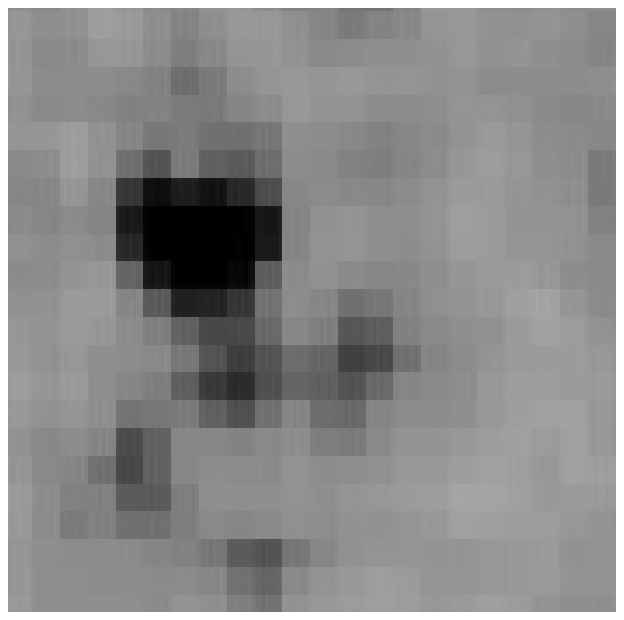}
\end{center}
\caption{Results for PKS1932-464.  Left -- Extracted
  [O\textsc{iii}]5007\AA\ flux from our VIMOS observations, in $\rm
  ergs\,s^{-1}\,cm^{-2}$. Centre -- [O\textsc{ii}]+continuum image of PKS1932-46
  \citep[from][]{vm98} illustrating the field of view of the VIMOS IFU data
  and object identifications. Right -- Spitzer MIPS images of the
  field (top: 70$\mu m$, bottom: 24$\mu m$).  PKS1932-464
  is undetected at the confusion limit at 70$\mu m$, and has a 24$\mu
  m$ flux of 2.6mJy.  The companion has 70$\mu m$ and 24$\mu m$ fluxes
  of 48.3 and 4.4mJy respectively.
\label{Fig: 1932}}
\end{figure*}

PKS1932-46 is an FRII radio galaxy at $z=0.231$.
Fig.~\ref{Fig: 1932} displays the  extracted [O\textsc{iii}]5007\AA\,
emission line flux from our IFS data, together with an [O\textsc{ii}]3727\AA+continuum
image of the field.  
We observe previously undetected line emitting material with a
very knotty morphology.  The knots seem to lie preferentially around
the cocoon edges rather than directly on the jet axis.  Interestingly, several knots lie at
large (projected) distances from the radio source axis (up to $\sim 20$kpc), well
outside the AGN ionization cone. While the emission to the west is
consistent with AGN photoionization, these knots extending N/S from
the host galaxy display a lower ionization state, placing them closer
to the maximum starburst track of \citet{kewley01}. We concur with
the interpretation of \citet{vm05}, i.e. that the EELR contains a
series of compact, star-forming objects, plausibly associated with the
tidal debris of a merging system.  

Star formation is not limited to the EELR 
knots, but is also clearly ongoing in the spiral companion galaxy to
the north ($z=0.229$), as can be seen from the relative luminosity of
the sources in our MIPS data.  In fact, PKS1932-464 is {\it underluminous}
in the mid-IR cf.\ $L_{[OIII]}$ compared to other, similar radio 
sources \citep{dicken07}, suggesting that on-going star formation is
restricted to the EELR and the galaxy itself contains relatively little  starburst-heated dust.   Together, our IFS and Spitzer data imply that
PKS1932-46 lies in a fairly disturbed environment, and that it may be
subject to the effects of both recent and incipient mergers/interactions.

\begin{figure*}
\vspace{3. in}
\begin{center}
\includegraphics{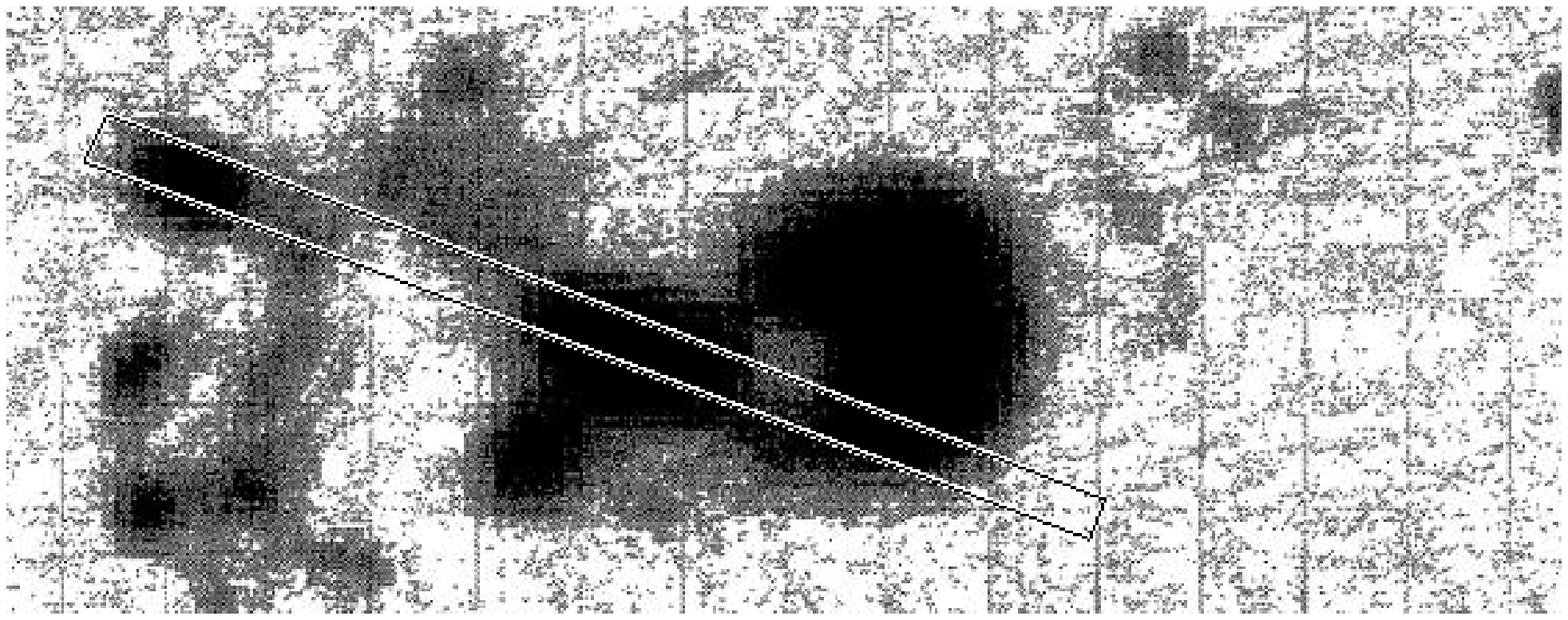}
\includegraphics{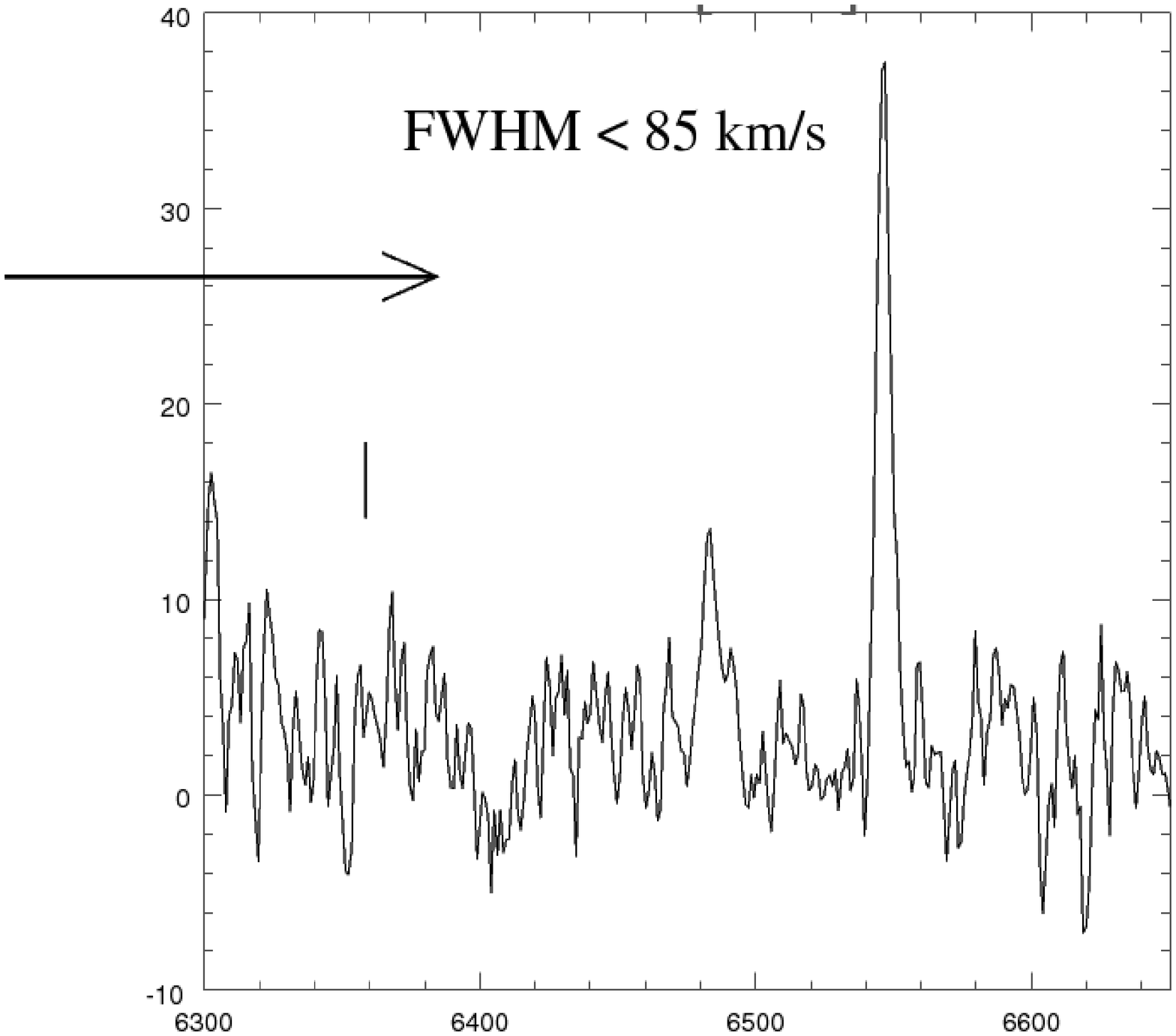}
\includegraphics{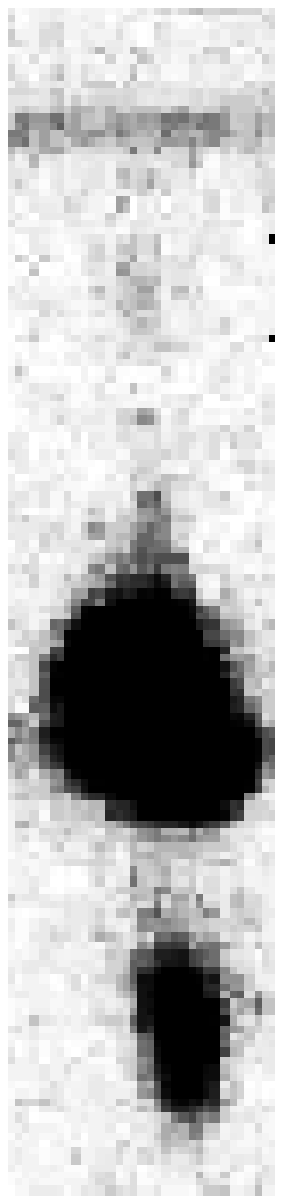}
\includegraphics{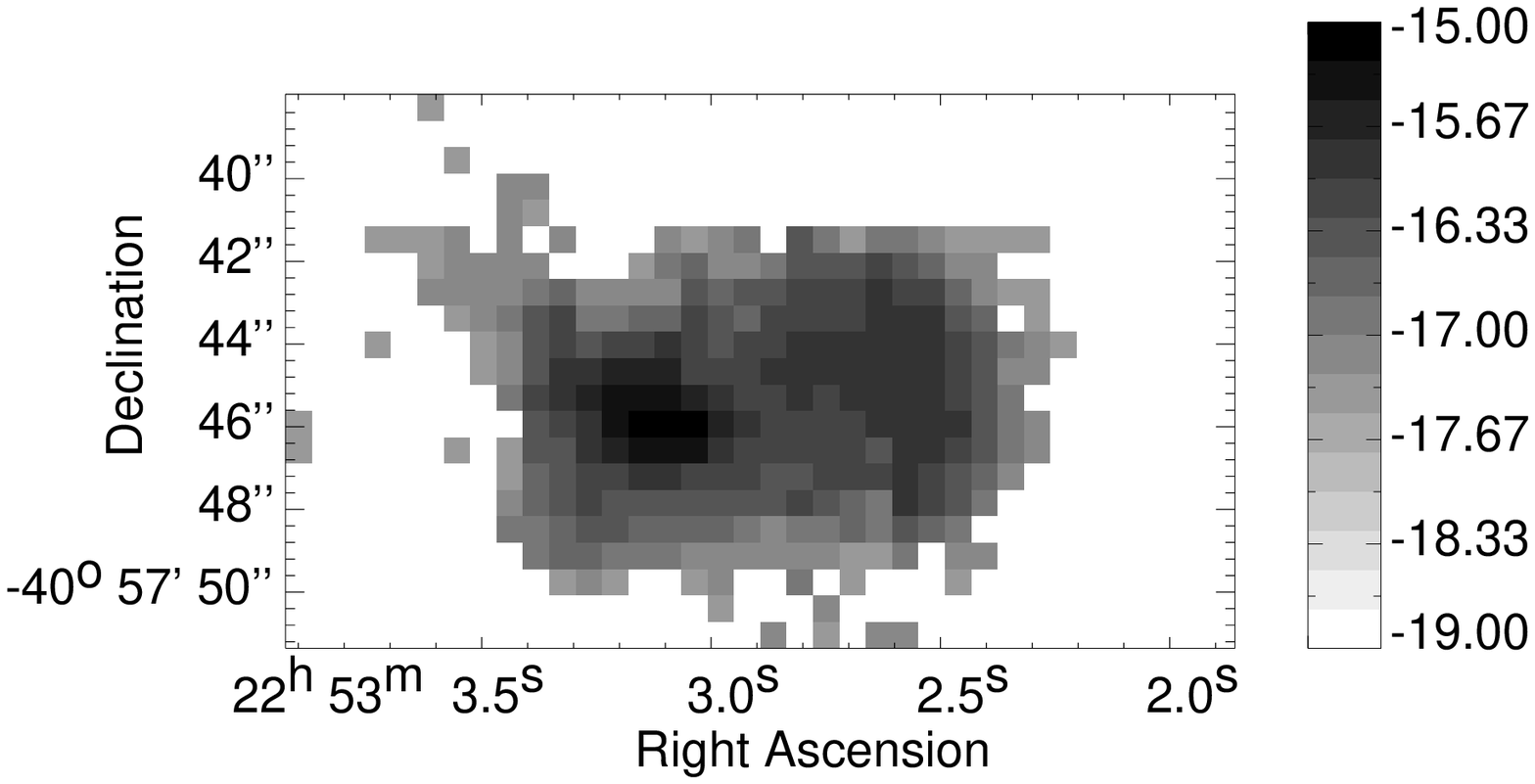}
\includegraphics{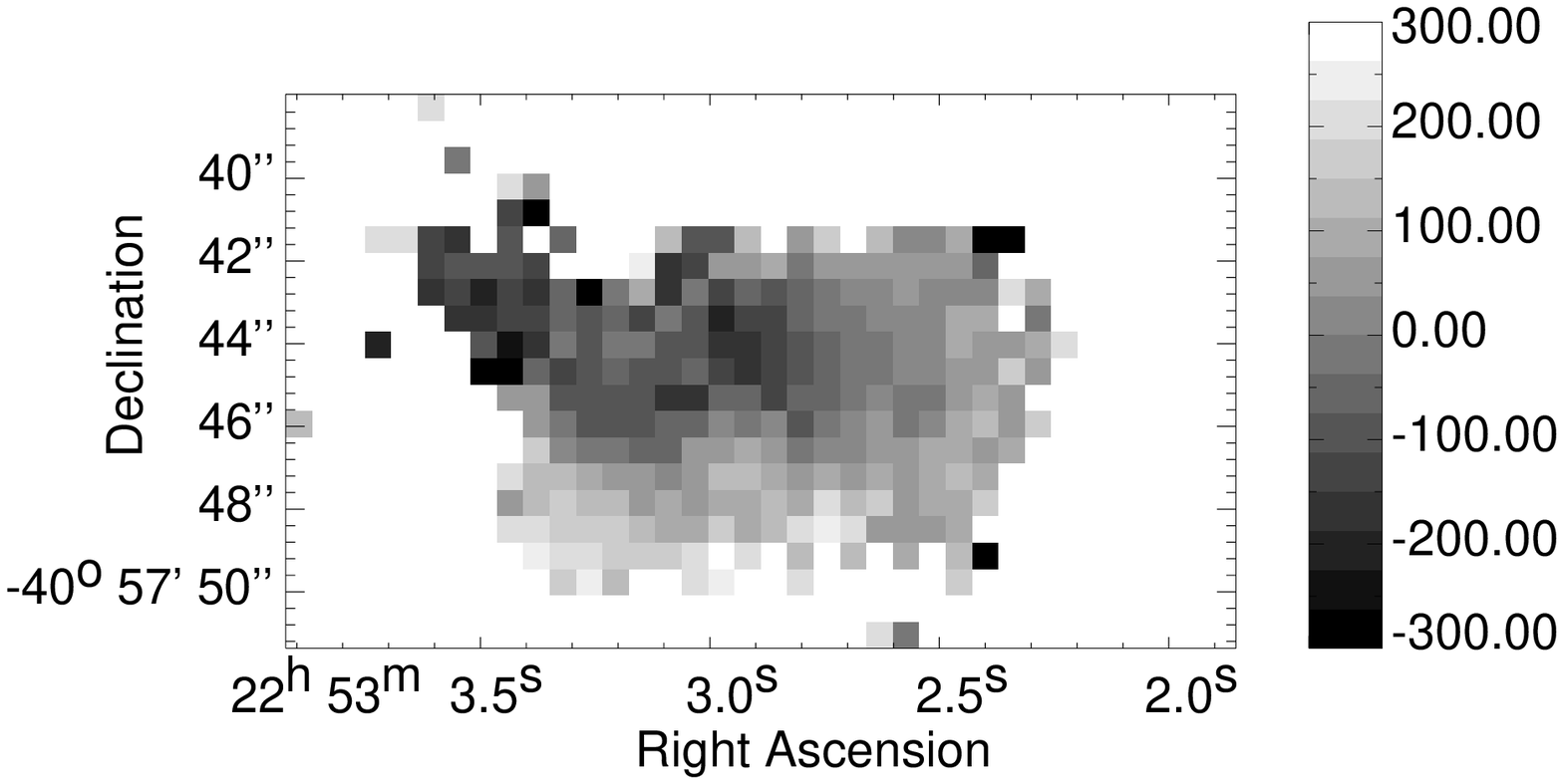}
\end{center}
\caption{Top row: VIMOS IFS data for PKS2250-41. Left -- extracted [O\textsc{iii}]5007\AA\ flux for PKS2250-41, in $\rm
  ergs\,s^{-1}\,cm^{-2}$; right -- velocity shift of the 
  [O\textsc{iii}]5007\AA\ emission relative to a rest-frame redshift of
  $z=0.308$ ($\rm km\,s^{-1}$). Bottom row: FORS1 spectroscopy of
  PKS2250-41.  Left -- [O\textsc{iii}]5007\AA+continuum image of
  PKS2250-41 \citep{clark97} illustrating the location of the
  $1^{\prime\prime}$ slit, scaled and aligned to match the
  [O\textsc{iii}]5007\AA\ IFS image above. Right -- 2-d spectrum displaying the
  [O\textsc{iii}]5007\AA\ emission line (east at top, west at bottom), and the 1-d spectrum of the diffuse
  emission between the radio source and the
  companion (indicated by arrow), displaying the
  [O\textsc{iii}]4959+5007 doublet and the 
  expected location of H$\beta$.
\label{Fig: 2250}}
\end{figure*}

PKS2250-41 (Fig.~\ref{Fig: 2250}) is an FRII radio galaxy at
$z=0.308$, and, like PKS1932-46, it also displays an EELR comparable in
size to the radio source. This EELR has been well-studied in the past
\citep{tad94,clark97,vm99,tilak05}, and the main emission line
features are knotty emission regions extending to the east, and a
luminous arc structure to the west, coincident with the western radio
lobe.  A detailed discussion of our IFS data in the context of EELR
ionization state, jet-cloud interactions and the kinematics of the
emission line arc can be found in \citet{inskip07}; here, we present a
few key results. The most extreme gas kinematics are associated with
the western radio jet and hotspot.  Away from the radio source axis, the more distant emission
displays unresolved line widths. We also see a clear rotation
profile across the EELR, confirmed in our FORS1 spectrum of
PKS2250-41.  The observed velocity curve is very
smooth, suggestive of a 20kpc rotating structure, and seems
unaffected by the more extreme kinematics in the central regions of 
the galaxy. 
The galaxy to the NE is also included in our FORS1 spectrum, and we
detect low S/N Balmer absorption features at a similar redshift to
PKS2250-41.  
The faint, diffuse band of emission  extending north-eastwards 
between the host galaxy and the companion
galaxy (resolved out in the higher
resolution emission line images of \citet{tilak05} but visible in the
\citet{clark97} data) displays unresolved line widths and appears to connect the two
systems, suggesting an ongoing interaction.

\section{Concluding discussion}

The expanding radio source strongly influences 
the properties of the surrounding EELR.  In addition to the usual
signs of interactions between EELR and the radio source, for
PKS1932-46, the EELR morphology appears to trace the edges of the
western radio lobe.  However, these two EELRs are far more extreme
than those typically observed around sources at these redshifts.
Both PKS1932-464 and PKS2250-41 display a number of features
suggestive of galaxy interactions: extensive gaseous halos, large-scale rotating structures
and close companion galaxies at similar
redshifts. In the case of
PKS1932-46, the companion galaxy is actively forming stars, as are
many of the knots in the EELR.  We suggest that both sources lie in rich, merger environments, and
that a large proportion of the EELR material may originate from the
tidal debris of previous interactions -- which may, plausibly, also
have been responsible for triggering the current radio source activity.




\end{document}